# Investigating Awareness and usage of Electronic Resources by the Library Users of Selected Colleges of Solapur University

Patel Adam Burhansab  
*Annamalai University*, pateladam.lib@gmail.com

Dr. M.Sadik Batcha  
*Annamalai University*, msbau@rediffmail.com

Muneer Ahmad  
*Annamalai University*, muneerbangroo@gmail.com

# Investigating Awareness and usage of Electronic Resources by the Library Users of Selected Colleges of Solapur University

Patel Adam Burhansab[1] Dr. M Sadik Batcha[2] Muneer Ahmad[3]

[1]*Research Scholar, Department of Library and Information Science, Annamalai University, Annamalai nagar, pateladam.lib@@gmail.com*

[2]*Research Supervisor and Mentor*, *Professor and Librarian, Annamalai University, Annamalai nagar, msbau@rediffmail.com*

[3]*Research Scholar, Department of Library and Information Science, Annamalai University, Annamalai nagar, muneerbangroo@gmail.com*

**Abstract**

The study investigated the use of electronic resources/information by library users in selected colleges of Solapur University. Specifically, to investigate the awareness and level of use of electronic resources; perceived reliance, benefits and impact of use of electronic resources on the research activities. The research design for the study was a survey. Questionnaire schedule was used to collect data from 1022 library users from selected colleges of Solapur University. The result revealed that preponderance of users from aided 33.51% Self financing 26.10% and Education colleges 43.24 % preferred to visit the Library once in three days. While analyzing the entire college libraries regarding the frequency of visit, users gave first preference to once in three days i.e. 27.2%. College wise analysis reveals that mainstream of users from Aided Colleges 38%, Self financing Colleges 28.3%, Engineering Colleges 43%, Education colleges 53.2% and Pharmacy Colleges 23.4% are spending their time 1-2 hrs in libraries and 40.8% visit college libraries to issue and return books and in the device usage (33.9%) of users ranked mobile phone as the second device for accessing the e-resources. It is noticeable that 24.7% of users acknowledge from Aided Colleges know about the services of the Library from website but in Self Finance Colleges 8.5% and Pharmacy College 56.4% users aware about the services from the friends it is observed that most of the web technologies are not used by the mainstream of the users due to lack of awareness, training etc.

## 1. Introduction

Electronic information resources have become essential in the provision of knowledge in research and development. Studies suggest that, scientists and development experts use

Electronic Resources (e-resources) to access, communicate and support Research and Development activities (Kumar & Singh, 2011). Among other things, Electronic Resources often supplement print resources with added advantages including the elimination of geographical barriers, provision of up-to-date information; convenience and the potential to explore additional resources or related content (Dadzie, 2005; Sharma, 2009; Renwidk, 2005; Agyei & Fiankor, 2017). E-resources are contributing immensely to information seeking and retrieval serving as a complement to print resources in traditional library settings (Tsakonas, & Papatheodorou, 2006) and a valuable alternative to print resources in academic and research domains Electronic Resources are those information sources which are accessible only via computer access, and may include the use of a microcomputer, a mainframe, or other types of computers. They can be locally mounted or accessed remotely via the internet (Chimni, 2018). (Graham, 2003) define electronic resources as that group of information sources explored through modern information and communication technology (ICT) devices. They can be stored in cyberspace and can be accessed simultaneously from different points by several people at the same time. The evolution of the concept and mode of presentation of information has introduced multimedia which involves the use of audio and visual forms and presents information in digital form/database often referred to as electronic form or electronic database (Chimni, 2018). E-resources may be delivered on Compact Disc Read-Only Memory (CDROM), on tape, via the internet (Chimni, 2018). The term electronic resources has broadly been defined as, information made available through a computer, which can either serve as bibliographic guides to other sources or as cited references in their own right (Graham,2003). Thus, documents which appear in electronic format and made available to users through any computer based information retrieval system are classified as electronic resources. Due to the fact that computers are the medium of access to e-resources, the internet is the most widely used channel (using search engines such as Google, Yahoo and Alta Vista) and then there are offline databases in Compact Disk/Digital Video Disc (CD/DVD) formats accessible without internet (Swain & Panda, 2009). Electronic Information Resources like print resources must sometimes be acquired, organised, stored and disseminated hence the need for libraries to provide support.

**2. Objectives**
- To measure users' satisfaction levels regarding electronic resources and services being offered in their libraries.

- To know users' opinions about user education/information literacy related to electronic resources and services.
- To identify problems faced by users in using electronic resources and services.

## 3. Research Methodology

### 3.1. Sampling and Questionnaire Design

A survey was conducted in 26 colleges of Solapur University based in Maharashtra, India. A structured questionnaire was designed. Library users within each institute were randomly. A total of one thousand and twenty two (1022) library users were randomly selected from the 26 institutes. The questionnaire consisted of various questions in the following categories: Demographics, use of e-resources, access to e-resources., competencies and training, challenges and benefits of e-resources among others. There were also two open-ended question that asked respondents about the impediment and improvement to e-resources in their institutes' libraries. In all 1050 questionnaires were distributed to the library users but 1022 were received. The response rate was 97 per cent. Statistical package for social sciences (SPSS version 16) was used to analyse the data.

### 3.2. Study Population

Out of 114 affiliated colleges to Solapur University, 26 colleges (including Aided Self, Financing Colleges, Engineering Colleges, Education Colleges & Pharmacy Colleges are selected for the study. The study population constituted library users of 26 college Libraries affiliated to Solapur University, Solapur.

### 3.3. Sample Selection

Sample is selected on the basis of confidence level 99% and confidence interval 5%. Following are the samples of the study of various colleges affiliated to Solapur University, Solapur. The sample and total population taken for the study are shown in Tabulation (Table 3.3)

Table 3.3: Representation of Respondents from Different Categories of colleges

| S.No | Category of the Colleges | Male | Female | Total |
|---|---|---|---|---|
| 1. | Aided College | 245 | 140 | 385 |
| 2. | Self Financing College | 174 | 144 | 318 |
| 3. | Engineering Colleges | 67 | 47 | 114 |
| 4. | Education Colleges | 66 | 45 | 111 |
| 5. | Pharmacy College | 59 | 35 | 94 |
| | **Total** | **611** | **411** | **1022** |

## 4. Results and Discussions

College wise analysis displays that preponderance of users from aided 33.51% Self financing 26.10% and Education colleges 43.24 % preferred to visit the Library once in three days. Majority of the users from Pharmacy 36.17% have given priority on daily frequency and another 21.53% visit library weekly; apart from this, About 32.46% users never visit the library however 23.58% users visit library on daily wise. In the entire colleges' monthly-wise visitor are found the least in numbers.

**Table 4.1: Frequency of visit in the Library (Colleges wise)**

| Frequency | Aided Colleges | | Self Financing Colleges | | Engineering Colleges | | Education Colleges | | Pharmacy Colleges | | Total | % |
|---|---|---|---|---|---|---|---|---|---|---|---|---|
| Daily | 101 | 26.23% | 64 | 20.13% | 16 | 14.04% | 26 | 23.42% | 34 | 19.77% | 241 | 48.69% |
| Once in 3 days | 129 | 33.51% | 83 | 26.10% | 8 | 7.02% | 48 | 43.24% | 10 | 5.81% | 83 | 16.77% |
| Weekly | 54 | 14.03% | 71 | 22.33% | 45 | 39.47% | 22 | 19.82% | 28 | 16.28% | 71 | 14.34% |
| Monthly | 27 | 7.01% | 34 | 10.69% | 8 | 7.02% | 7 | 6.31% | 34 | 19.77% | 34 | 6.87% |
| Never | 74 | 19.22% | 66 | 20.75% | 37 | 32.46% | 8 | 7.21% | 66 | 38.37% | 66 | 13.33% |
| Total | 385 | 100.00% | 318 | 100.00% | 114 | 100.00% | 111 | 100.00% | 172 | 100.00% | 495 | 100.00% |
| Chi.square 406.9 | | | | | | | | | | | P.Value =0.000 | |

While analyzing the entire college libraries regarding the frequency of visit, users gave first preference to once in three days i.e. 27.2%. Apart from this, a good number of users visiting the library on daily 23.58% and weekly 21.53%.basis. There is a slighter difference in the ease of occasional visitor which is observed to 19.08%. When compared to weekly visitors, it is 21.5%, only 8.6% of users are visiting library on monthly-wise which is noted a minor group. Instead of ignoring the minor group, authorities should encourage the users in visiting library on daily wise or at least once in three days. It is revealed that major groups of users visiting the library once in three days or daily wise. It brings out the focus on the importance and influence of library in the academic field. Library is the heart of any organization for collecting and disseminating of information. In addition chi Square value is 406.9 and p.value zero shows a significant variation in the frequency of visiting library by the users in various affiliated college at Solapur University, Solapur and it is at 1% level.

**Table 4.2: Time spent during a visit in the library (College wise)**

| Frequency | Aided Colleges | | Self Financing Colleges | | Engineering Colleges | | Education Colleges | | Pharmacy Colleges | | Total | % |
|---|---|---|---|---|---|---|---|---|---|---|---|---|
| Less than 1 Hr. | 84 | 21.82% | 149 | 46.86% | 42 | 36.84% | 29 | 26.13% | 56 | 59.57% | 56 | 59.57% |
| 2 hrs. | 146 | 37.92% | 90 | 28.30% | 49 | 42.98% | 59 | 53.15% | 22 | 23.40% | 22 | 23.40% |
| 2-3 hrs | 115 | 29.87% | 47 | 14.78% | 12 | 10.53% | 15 | 13.51% | 9 | 9.57% | 9 | 9.57% |
| More than 3 hrs. | 40 | 10.39% | 32 | 10.06% | 11 | 9.65% | 8 | 7.21% | 7 | 7.45% | 7 | 7.45% |
| Total | 385 | 100.00% | 318 | 100.00% | 114 | 100.00% | 111 | 100.00% | 94 | 100.00% | 94 | 100.00% |
| Chi. Square 360.3 | | | | | | | | | | | | P.value=0.000 |

User responses towards the total time spend in their college libraries are given in table 4.2. College wise analysis reveals that mainstream of users from Aided Colleges 38%, Self financing Colleges 28.3%, Engineering Colleges 43%, Education colleges 53.2% and Pharmacy Colleges 23.4% are spending their time 1-2 hrs in libraries. However higher numbers of users from Aided Colleges (29.8%), Self financing Colleges (14.8%), Engineering Colleges (10.5%), Education Colleges (13.5%) and Pharmacy Colleges (9.6 %) spend 2-3 hrs in a visit. The highest percentage at users spending more than 3 hours in a visit is from aided colleges is 32.46%, Self Financing colleges is 32.76%, Self Financing colleges is 32.76%, Engineering colleges is 14.28%, Education colleges is 10.38% and pharmacy colleges is 9.09%. They are spending more than three hours in a visit.

Overall analysis found that, frequency of use in the college libraries by the users is not satisfactory even though the libraries have rich collection of information which is essential for their academic work. Thus authorities should take vital actions to create interest among the users to use the library and its resources. Since the frequency of library use is fewer among aided and self financing and engineering hence more measures should be taken by the colleges to improve their collection and services to attract more users. It is also advocates that proper orientation should be given to users for effective utilization of their libraries.

**Table 4.3: Purpose of Library Visits**

| Purpose | Preferences | | | | | | | | | | | | Total | Rank |
| | 1 | | 2 | | 3 | | 4 | | 5 | | 6 | | | |
|---|---|---|---|---|---|---|---|---|---|---|---|---|---|---|
| Electronic Sources | 99 | 9.7 % | 95 | 9.3 % | 141 | 13.8% | 129 | 12.6% | 200 | 19.6% | 358 | 35.0% | 1022 | 6 |

| | | | | | | | | | | | | | |
|---|---|---|---|---|---|---|---|---|---|---|---|---|---|
| Issues Return | 417 | 40.8 % | 154 | 15.1% | 140 | 13.7% | 164 | 16.0% | 75 | 7.3% | 72 | 7.0% | 1022 | 1 |
| Reading Periodical | 69 | 6.8 % | 156 | 15.3 % | 93 | 9.1 % | 212 | 20.7% | 306 | 29.9% | 186 | 18.2% | 1022 | 5 |
| References | 135 | 13.2 % | 182 | 17.8% | 201 | 19.7% | 283 | 27.7% | 133 | 13.0% | 88 | 8.6% | 1022 | 4 |
| News paper Reading | 152 | 14.9 % | 137 | 13.4% | 222 | 21.7% | 180 | 17.6% | 127 | 12.4% | 204 | 20.0% | 1022 | 3 |
| Data Collection | 137 | 13.4% | 295 | 28.9% | 192 | 18.8% | 120 | 11.7% | 130 | 12.7% | 148 | 14.5% | 1022 | 2 |

Libraries play a crucial role to accomplish the visions and missions of their users. The purpose of visiting library may vary from users to users. Some of them are interested in newspaper reading and some others searching e-resources for preparation of projects, assignments, examinations etc. which depends upon their flavors. If the libraries make available the resources and services in order to cater the needs of the users with limited time using latest technologies, it will support to raise the number of visitors.

Overall analysis to the responds to the query regarding the purpose of visit of their Library reveals that the first and foremost purpose of visit by the users community of the various selected colleges libraries is the issue and return of books (40.8%). It shows the importance and influence of printed materials for reading among the users community. Nowadays project preparation and assignments are the part of each course, thus the users (28.9%) given second ranking to the data collection for writing assignment on specific information in their subject fields. News paper reading shows the users' enthusiasms to identify the recent general knowledge have been chosen in the third position. About 21.7% users are supported news paper reading.

**Table 4.4: Devices used for access IT resources (College-wise)**

| Devices | Preferences | | | | | | | | | | Total | |
|---|---|---|---|---|---|---|---|---|---|---|---|---|
| | 1 | | 2 | | 3 | | 4 | | 5 | | | |
| Computer of the Libraries | 451 | 44.1 % | 134 | 13.1 % | 209 | 20.5 % | 114 | 11.2 % | 114 | 11.2 % | 1022 | 100 % |
| Computer of the other Libraries | 103 | 10.1 % | 178 | 17.4 % | 214 | 20.9 % | 400 | 39.1 % | 127 | 12.4 % | 1022 | 100 % |

| | | | | | | | | | | | |
|---|---|---|---|---|---|---|---|---|---|---|---|
| Laptop/Note Book | 176 | 17.2 % | 269 | 26.3 % | 281 | 27.5 % | 199 | 19.5 % | 97 | 9.5 % | 1022 | 100 % |
| Mobile Phone | 130 | 12.7 % | 346 | 33.9 % | 273 | 26.7 % | 176 | 17.2 % | 97 | 9.5 % | 1022 | 100 % |
| I Pad Tablets | 90 | 8.8 % | 77 | 7.5 % | 116 | 11.4 % | 168 | 16.4 % | 571 | 55.9 % | 1022 | 100 % |
| Total | 950 | 18.6 % | 1004 | 19.6 % | 1093 | 21.4 % | 1057 | 20.7 % | 1006 | 19.7 % | 5110 | 100 % |

While ranking the device used for retrieving the Library resources it is explicable that the best part of the users use the computers of college libraries. For accessing the IT enabled services. Now the shape, size and facilities of the mobile phones are changing day by day and it is inevitable from the daily life. Most of the users are preferred mobile phone for their network access.

In the device usage (33.9%) of users ranked mobile phone as the second device for accessing the e-resources. It is evident that 27.5% of users use laptop/ notebook. For accessing the IT resources of Libraries and ranked it in the third position. Now technologies are changing and the chances for using the laptop /notebook are increasing. Within one or two years it will come in the First or in the Second position for retrieving the e-resources. It is noticeable that 39.2% users are using the devices of other Libraries and given fourth preference. It may be due to the lack of technologies in some Libraries. I pad/ Tablets are now common among the users, but due to some restriction it is not encouraged in the computer.

**Table 4.5: Acknowledged about IT Services in the Libraries (College-wise)**

| Colleges | Aided Colleges | | Self Finance College | | Engineering Colleges | | Education Colleges | | Pharmacy Colleges | | Total | |
|---|---|---|---|---|---|---|---|---|---|---|---|---|
| Staff | 65 | 16.9 % | 43 | 13.5 % | 46 | 40.4 % | 44 | 39.6 % | 15 | 16.0 % | 213 | 20.8 % |
| Friends | 82 | 21.3 % | 124 | 39.0 % | 19 | 16.7 % | 42 | 37.8 % | 53 | 56.4 % | 320 | 31.3 % |
| Websites | 50 | 24.7 % | 27 | 8.5 % | 15 | 13.2 % | 8 | 7.2 % | 7 | 7.4 % | 152 | 14.9 % |

| Teachers | 72 | 28.1 % | 89 | 28.0 % | 8 | 7.0 % | 10 | 9.0 % | 13 | 13.8 % | 228 | 22.3 % |
|---|---|---|---|---|---|---|---|---|---|---|---|---|
| Orientation | 198 | 9.1 % | 35 | 11.0 % | 26 | 22.8 % | 7 | 6.3 % | 6 | 6.4 % | 109 | 10.7 % |
| Total | 385 | 100 % | 318 | 100 % | 114 | 100 % | 111 | 100 % | 94 | 100 % | 1022 | 100 % |
| Chi Square = 715.6 | | | | | | | | | | | P Value = 0.000 | |

Analysis of the table 4.5 shows that 24.7% of users from Aided Colleges know about the services of the Library from website but in Self Finance Colleges 8.5% and Pharmacy College 56.4% users aware about the services from the friends. In Engineering Colleges (40.4%) and colleges (39.6%) staff played a key role to spend the awareness about the resources of the Library. Apart from this (22.8%) of users from Engineering colleges attains knowledge through orientation programmers and (39.8%) from Education Colleges through friends. Chi-square value 715.6 and p.value zero which is less than 0.01 indicate that there is a significant difference among the users in Libraries regarding the sources thorough which the users know about the IT services provided by the Library.

Overall analysis of the whole colleges show that (31.3%) of users got information regarding library IT services through Friends Teachers (22.3%) and Staff (20.8%) also provide valuable information regarding the IT based library services to their users. Only (14.9%) study knows IT resources through websites. According to the users orientation programmer (10.7%) are not much effective in conveying the knowledge about IT services. This approach may be due to inappropriate usage of funds granted to them for the upliftment of users or may be due to lack of funds available. The most important reason is that the users attending such Library orientation programs may be few or may be unaware of the program conducted by the Libraries. The results give a clear picture that teacher, staff, friends and library orientation programs can play a key role to know about the resources available in the libraries and its maximum utilization to the users. In the technology revolutionary world the colleges' libraries website also perform a vital role for providing various online services offered by the library. But regrettably, majority of the college libraries are not providing services through website except Aided Colleges Self financing colleges.

**Table 4.6: Training in IT based Resources and Services**

| Colleges | | Aided Colleges | | Self Finance College | | Engineering Colleges | | Education Colleges | | Pharmacy Colleges | | Total | |
|---|---|---|---|---|---|---|---|---|---|---|---|---|---|
| Opinion | Yes | 65 | 16.9 % | 19 | 6.0 % | 26 | 22.8 % | 12 | 10.8 % | 13 | 13.8 % | 135 | 13.2 % |
| | No | 320 | 83.1 % | 299 | 94.0 % | 88 | 77.2 % | 99 | 89.2 % | 81 | 86.2 % | 887 | 86.8 % |
| Total | | 385 | 100 % | 318 | 100 % | 114 | 100 % | 111 | 100 % | 94 | 100 % | 1022 | 100 % |

Chi Square = 76.9     P Value = 0.000

It is noticeable from the table 4.6 that majority of users from Aided Colleges (83.1%) Self Financing Colleges (94.0%) Engineering Colleges (77.2%) Education Colleges (89.2%) and Pharmacy Colleges (86.2%) users are aware of the training programs compared to the users in other colleges. Users from Self Financing Colleges (94.0%) and Education Colleges (89.2%) mostly claim that they are not aware of such training programs in Libraries. The main reason for such unawareness regarding the IT training in using library facilities may be due to less publicity regarding these services. In this era of information technology only proper training of using Library IT facilities can help the users to exploit the service positively in accordance with their needs to improve their quality of their work.

The analysis of data shows that much higher percentage of users (86.6%) claim that they are not getting any sort of training programmes only (13.2%) of users admits that they are aware of such training and also use them to meet their knowledge demands. Thus there arises a higher need for IT training programs workshops in the college libraries for the maximum utilization of the library services. More publicity regarding this should be encouraged. The college libraries conducting various training program for the users but unfortunately the results reveals that the success rate of the training is less. The chi-square value is 79.9 and p-value is 0.000 which is less than 0.01. Thus it is clear that there is a significant difference among the users in the different college Libraries regarding the opinion on IT based training conducted by the library.

**Table 4.7: Web Technology Usage**

| Purpose | Daily | | Weekly | | Monthly | | Once in a While | | Never | | Total | Remark |
|---|---|---|---|---|---|---|---|---|---|---|---|---|
| Internet Surfing | 380 | 37.2 % | 260 | 25.4% | 106 | 10.4% | 94 | 9.2% | 182 | 17.8% | 1022 | 2(Daily) |
| Search Engine | 376 | 36.8 % | 322 | 31.5% | 95 | 9.3% | 92 | 9.0% | 137 | 13.4% | 1022 | |
| E-mails | 433 | 42.4% | 230 | 22.5% | 141 | 13.8% | 120 | 11.7% | 98 | 9.6% | 1022 | |
| Blogs | 134 | 13.1% | 204 | 20.0% | 193 | 18.9% | 120 | 11.7% | 371 | 36.3% | 1022 | |
| E-Resources Search | 130 | 12.7% | 276 | 27.0% | 192 | 18.8% | 140 | 13.7% | 284 | 27.8% | 1022 | |
| Downloading of E-Documents | 181 | 17.7% | 251 | 24.6% | 124 | 12.1% | 160 | 15.7% | 306 | 29.9% | 1022 | |
| Rss Feeds (Journal) | 114 | 11.2% | 185 | 18.1% | 113 | 11.1% | 94 | 9.2% | 516 | 50.5% | 1022 | |
| Mobile Alert Services | 157 | 15.4% | 248 | 24.3% | 136 | 13.3% | 103 | 10.1% | 378 | 37.0% | 1022 | |
| Wikis | 194 | 19.0% | 312 | 30.5% | 128 | 12.5% | 110 | 10.8% | 278 | 27.2% | 1022 | |
| Web conferencing | 90 | 8.8% | 160 | 15.7% | 126 | 12.3% | 184 | 18.0% | 462 | 45.2% | 1022 | |
| Video Sharing | 93 | 9.1% | 248 | 24.3% | 146 | 14.3% | 220 | 21.5% | 315 | 30.8% | 1022 | |
| Photo Sharing | 135 | 13.2% | 260 | 25.4% | 170 | 16.6% | 191 | 18.7% | 266 | 26.0% | 1022 | |
| Instant Messaging | 180 | 17.6% | 221 | 21.6% | 160 | 15.7% | 165 | 16.1% | 296 | 29.0% | 1022 | |
| E-Learning Management System | 112 | 11.0% | 163 | 15.9% | 151 | 14.8% | 169 | 16.5% | 427 | 41.8% | 1022 | |
| Digital Library Services | 97 | 9.5% | 138 | 13.5% | 143 | 14.0% | 137 | 13.4% | 507 | 49.6% | 1022 | |
| Content Management System | 69 | 6.8% | 155 | 15.2% | 126 | 12.3% | 143 | 14.0% | 529 | 51.8% | 1022 | |

Analyzing the entire college web technology usages, it is observed from the table 4.7 that most of the web technologies are not used by the mainstream of the users due to lack of awareness, training etc. Accurate and appropriate training should be conducted by the colleges Libraries according to the necessities of the users. Systematic training will inevitably help the user for maximum utilization of e-resources of the library. The leading web technologies such as internet surfing, emails, search engines, wikis, photo sharing etc. are used by the great number of users on daily and weekly basis of frequency. On the other hand majority of the web technologies are never used by the great number of users.

**5. Conclusion and Recommendations**

This study investigated the awareness and use of electronic resources among library users of the selected colleges of Solapur University. The study revealed that e-resources awareness and use is very common among library users. Library users largely depend on e-resources to search, retrieve and communicate research making e-resources an essential party to finding reliable, timely and relevant information. The use of electronic resources had also improved their research output. The most used electronic resources were e-databases and e-research reports and e-journals. However, considering the high investments made in acquiring these resources; perhaps, improvements in the infrastructure would lead to increased use of these resources by research scientists. In addition, continuous training on information literacy skills, emphasizing on electronic information retrieval would go a long way to enhance the effective use of electronic resources and the numerous benefits of electronic resources would be appreciated. User training is essential particularly for library users to enable them search independently unlike in traditional library setting where librarians could retrieve information for users. Training on advanced search strategies controlled vocabulary and general internet use for scholarly and academic purposes should be organised to make electronic search processes much easier. Finally, computer knowledge and information retrieval skills must be frequently enhanced to keep up with evolving information communication technology.